\documentclass{elsart}

 \usepackage{graphics}
 \usepackage{graphicx}
 \usepackage{epsfig}

\usepackage{amssymb}

\newcounter{bla}
\newenvironment{refnummer}{%
\list{[\arabic{bla}]}%
{\usecounter{bla}%
 \setlength{\itemindent}{0pt}%
 \setlength{\topsep}{0pt}%
 \setlength{\itemsep}{0pt}%
 \setlength{\labelsep}{2pt}%
 \setlength{\listparindent}{0pt}%
 \settowidth{\labelwidth}{[9]}%
 \setlength{\leftmargin}{\labelwidth}%
 \addtolength{\leftmargin}{\labelsep}%
 \setlength{\rightmargin}{0pt}}}
 {\endlist}

\begin{document}

\begin{frontmatter}


\title{LCG MCDB -- a Knowledgebase of Monte-Carlo Simulated Events.} 


\author[jinr]{S.~Belov} 
\author[sinp]{L.~Dudko} 
\author[ihep]{E.~Galkin} 
\author[ihep]{A.~Gusev} 
\author[cern]{W.~Pokorski} 
\author[cambridge,sinp]{A.~Sherstnev\corauthref{corts}} 

\address[jinr]{Joint Institute for Nuclear Research, Dubna, Moscow region, Russia, 141980} 
\address[sinp]{Scobeltsyn Institute of Nuclear Physics of Lomonosov Moscow State University, 
Moscow, Russia, 119992}
\address[ihep]{Institute For High Energy Physics, Protvino, Russia, 142281} 
\address[cern]{CERN/SFT, CH-1211 Geneva 23, Switzerland} 
\address[cambridge]{Cavendish Laboratory, University of Cambridge, CB3 0HE, UK} 

\corauth[corts]{Corresponding author. \textit{Email address}: \texttt{cherstn@hep.phy.cam.ac.uk}.}

\begin{abstract}
In this paper we report on LCG Monte-Carlo Data Base (MCDB) and software 
which has been developed to operate MCDB. The main purpose of the LCG MCDB 
project is to provide a storage and documentation system for sophisticated 
event samples simulated for the LHC collaborations by experts. In many cases, 
the modern Monte-Carlo simulation of physical processes requires expert 
knowledge in Monte-Carlo generators or significant amount of CPU time to 
produce the events. MCDB is a knowledgebase mainly dedicated to accumulate 
simulated events of this type. The main motivation behind LCG MCDB is to make 
the sophisticated MC event samples available for various physical groups. 
All the data from MCDB is accessible in several convenient ways. LCG MCDB 
is being developed within the CERN LCG Application Area Simulation project. 
\end{abstract}

%\begin{keyword}
%MCDB \sep Monte-Carlo Simulation \sep Monte-Carlo Generators \sep LHC \sep Knowledgebase
%\PACS 29.85.+c %Computer data analysis
%\PACS 01.50.hv %Computer software and software reviews
%\PACS 02.70.-c %Computational techniques; simulations
%\PACS 02.70.Uu %Applications of Monte-Carlo methods 
%\PACS 07.05.-t %Computers in experimental physics
%\PACS 07.05.Tp %Computer modeling and simulation
%\PACS 07.05.Wr %Computer interfaces

%%%%%%%%%%%%% Chosen PACS %%%%%%%%%%%%%
%\PACS 01.50.hv 
%\PACS 07.05.-t 
%\PACS 07.05.Tp 
%\PACS 07.05.Wr 

%\end{keyword}
\end{frontmatter}

%%%%%%%%%%%%%%%%%%%%%%%%%%%%%%%%%%%%%%%%%%%%%%%%%%%%%%%%%%%%%%%%%%%%%%%%%%%%%%%%%%%%%%%%%%%%%%%%
\section{Program Summary}
\label{progsum}

\begin{small}
\noindent
{\em Manuscript Title: LCG MCDB -- a Knowledgebase of Monte-Carlo Simulated Events.} \\
{\em Authors: S.~Belov, L.~Dudko, E.~Galkin, A.~Gusev, W.~Pokorski, A.~Sherstnev} \\
{\em Program Title:LCG Monte-Carlo Data Base} \\
{\em Journal Reference: }                                      \\
  %Leave blank, supplied by Elsevier.
{\em Catalogue identifier:}                                   \\
  %Leave blank, supplied by Elsevier.
{\em Licensing provisions: GPL}                                   \\
  %enter "none" if CPC non-profit use license is sufficient.
{\em Programming language: Perl}\\
{\em Computer: CPU: Intel Pentium 4, RAM: 1 Gb, HDD: 100 Gb} \\
  %Computer(s) for which program has been designed.
{\em Operating system: Scientific Linux CERN 3/4} \\
  %Operating system(s) for which program has been designed.
{\em RAM: 1073741824 bytes (1 Gb)} \\
  %RAM in bytes required to execute program with typical data.
{\em Number of processors used: 1}                              \\
  %If more than one processor.
{\em Supplementary material:}                                 \\
  % Fill in if necessary, otherwise leave out.
{\em Keywords: MCDB, Monte-Carlo Simulation, Monte-Carlo Generators, LHC, Knowledgebase.}\\
  % Please give some freely chosen keywords that we can use in a
  % cumulative keyword index.
{\em PACS: 01.50.hv, 07.05.-t, 07.05.Tp, 07.05.Wr}\\
  % see http://www.aip.org/pacs/pacs.html 
{\em Classification: 9 Databases, Data Compilation and Information Retrieval} \\
  %Classify using CPC Program Library Subject Index, see (
  % http://cpc.cs.qub.ac.uk/subjectIndex/SUBJECT_index.html)
  %e.g. 4.4 Feynman diagrams, 5 Computer Algebra.
{\em External routines/libraries: \\
perl $>=$ 5.8.5; Perl modules (DBD-mysql $>=$ 2.9004, File::Basename, GD::SecurityImage, 
\\ GD::SecurityImage::AC, Linux::Statistics, XML::LibXML $>$ 1.6, XML::SAX, \\
XML::NamespaceSupport); Apache HTTP Server $>=$ 2.0.59; 
mod\_auth\_external $>=$ 2.2.9; edg-utils-system RPM package; gd $>=$ 2.0.28; 
rpm package CASTOR-client $>=$ 2.1.2-4; arc-server (optional).
}
\\
{\em Subprograms used:}                                       \\
  %Fill in if necessary, otherwise leave out.
{\em Catalogue identifier of previous version:}               \\
  %Only required for a New Version summary, otherwise leave out.
{\em Journal reference of previous version:}                  \\
  %Only required for a New Version summary, otherwise leave out.
{\em Does the new version supersede the previous version?:}    \\
  %Only required for a New Version summary, otherwise leave out.
{\em Nature of problem:\\
%Describe the nature of the problem here.
Often, different groups of experimentalists prepare similar samples of particle 
collision events or turn to the same group of authors of Monte-Carlo~(MC) 
generators to prepare the events. 
For example, the same MC samples of Standard Model (SM) processes can be 
employed for the investigations either in the SM analyses (as a signal) or 
in searches for new phenomena in Beyond Standard Model analyses (as a background). 
If the samples are made available publicly and equipped with corresponding 
and comprehensive documentation, it can speed up cross checks of the samples 
themselves and physical models applied. Some event samples require a lot 
of computing resources for preparation. So, a central storage of the samples 
prevents possible waste of researcher time and computing resources, which can 
be used to prepare the same events many times. 
}
\\
{\em Solution method:\\
%Describe the method solution here.
Creation of a special knowledgebase (MCDB) designed to keep event samples 
for the LHC experimental and phenomenological community. The knowledgebase 
is realized as a separate web-server (mcdb.cern.ch). All event samples are 
kept on types at CERN. Documentation described the events is the main contents 
of MCDB. Users can browse the knowledgebase, read and comment articles 
(documentation), and download event samples. Authors can upload new event 
samples, create new articles, and edit own articles. 
}
\\
{\em Reasons for the new version:}\\
  %Only required for a New Version summary, otherwise leave out.
{\em Summary of revisions:}\\
  %Only required for a New Version summary, otherwise leave out.
{\em Restrictions:\\
  %Describe any restrictions on the complexity of the problem here.
The software is adopted to solve the problems, described in the article
and there are no any additional restrictions.
}
\\
{\em Unusual features:\\
  %Describe any unusual features of the program/problem here.
The software provides a framework to store and document large files with
flexibal authentication and authorization system. Different external storages with
large capacity can be used to keep the files. The WEB Content Management System
provides all of the necessary interfaces for the authors of the files, end-users 
and administrators.
}
\\
{\em Additional comments:}\\
  %Provide any additional comments here.
{\em Running time:\\
  %Give an indication of the typical running time here.
Real time operations.
}  
\\
{\em References:}
\begin{refnummer}
\item 
The main LCG MCDB server \verb|http://mcdb.cern.ch/|

\item 
P.~Bartalini, L.~Dudko, A.~Kryukov, I.~V.~Selyuzhenkov, A.~Sherstnev and A.~Vologdin, 
``LCG Monte-Carlo data base,'' 
[arXiv:hep-ph/0404241].

\item 
J.~P.~Baud, B.~Couturier, C.~Curran, J.~D.~Durand, E.~Knezo, S.~Occhetti and O.~Barring,
``CASTOR: status and evolution,''
[arXiv:cs.oh/0305047].

\end{refnummer}

\end{small}

\hspace{1pc}
{\bf LONG WRITE-UP}
%%%%%%%%%%%%%%%%%%%%%%%%%%%%%%%%%%%%%%%%%%%%%%%%%%%%%%%%%%%%%%%%%%%%%%%%%%%%%%%%%%%%%%%%%%%%%%%%
\section{Introduction}
\label{intro}

The LCG MCDB project~\cite{Dobbs:2004bu,Bartalini:2004nd} has been created 
to facilitate communication between experts/authors of Monte-Carlo (MC) 
generators and users of the programs in the LHC collaborations. 

The current version of LCG MCDB provides flexible infrastructure to share samples 
of events of particle collisions in accelerators prepared by the MC method 
(MC event samples) and keep the files in a reliable and convenient way. 
It has several interfaces, mainly Web-based, which help to carry out 
routine operations with stored samples by end-users and authors of the samples. 

LCG MCDB is particularly useful in tasks where the preparation of event 
samples requires specific knowledge of the Monte-Carlo codes/techniques 
applied, significant computing power, and/or constant interaction between 
end-users and the authors of the events. 
In many standard tasks events can be produced ``on the fly'' keeping just 
initial ``data-cards'', i.e. MC code parameter values which unambiguously 
define a concrete simulation run. But if the simulation time or exploitable 
resources become a significant factor, it would be worth considering the 
event sample as a candidate to keep in LCG MCDB. For instance, this situation 
can arise if we use such MC programs as ALPGEN~\cite{Mangano:2002ea}, 
CompHEP~\cite{Boos:2004kh}, GRACE~\cite{Yuasa:1999rg}, or MadGraph~\cite{Maltoni:2002qb}. 
Even MC generators as PYTHIA or HERWIG sometimes require the keeping of 
event files themselves. Examples of this sort happen in simulations of rare 
processes and/or with strong pre-selection cuts. 

The second motivation behind the project is to create a central database of 
MC events, where stored event samples are publicly available for all groups 
to use and/or validate. Often, different groups of experimentalists prepare 
similar event samples or turn to the same group of authors of MC generators 
in the simulation. For example, the same MC samples of Standard Model (SM) 
processes can be employed for the investigations either in the SM analyses 
(as a signal) or in searches for new phenomena in Beyond Standard Model analyses 
(as a background). Publicity of the event samples equipped with corresponding 
and comprehensive documentation can speed up cross checks of the samples 
themselves and physical models applied. It also prevents possible waste 
of researcher time and computing resources. 

The previous version~\cite{cms_mcdb_url} of MCDB was launched by the CMS 
collaboration in 2002. Several years of extensive use of the database 
have shown some limitations of a design applied in CMS MCDB. Storage of 
event files on AFS~\footnote{The Andrew File System (AFS) is a distributed 
networked file system developed by Carnegie Mellon University as part of the 
Andrew Project.} allows one to keep only small sized MC samples (basically 
parton level events prepared by Matrix Element tools), phonetic-based search 
turned out to be insufficient, and the database does not have simple tools 
to reuse information entered in MCDB earlier, such as physical parameters 
(masses, couplings, etc.), process information (name, PDF, particle content), 
generator information, etc. 
The new design of MCDB presented in this paper overcomes these problems and 
gives opportunities for further development of the idea. LCG MCDB is based 
on much more powerful, standardized and exportable software tools that are 
available to the LHC collaborations. Current migration of CMS physical groups 
from old CMS MCDB to the LCG framework gives us the motivation for the 
further development of the tool. 

In the next sections we describe the LCG MCDB design and ideas in more 
detail and briefly portray subsystems and modules of LCG MCDB. 
Section~\ref{howto} reports how end-users can use the software. A more 
detailed manual and installation instructions are available on the LCG 
MCDB server (~\cite{lcg_mcdb_url}, help section). 

At present, LCG MCDB is a stable software package and ready to use for 
the LHC community. A dedicated web server is deployed in~\cite{lcg_mcdb_url}. 

%%%%%%%%%%%%%%%%%%%%%%%%%%%%%%%%%%%%%%%%%%%%%%%%%%%%%%%%%%%%%%%%%%%%%%%%%%%%%%%%%%%%%%%%%%%%%%%%
\section{LCG MCDB as a knowledgebase} 
\label{concept} 

Knowledgebase is a special kind of database for knowledge management. It 
provides the means for the computerized collection, organization, and 
retrieval of knowledge~\cite{wikipedia_kb}. According to the definition 
one of the specific features of knowledgebase is that it keeps metadata or 
meta-information, i.e. information on data. Usually it is not possible 
to strictly distinguish between data and metadata, since the separation 
depends on situations where the data are exploited. In our concrete case 
we discriminate between events, as sets of particle 4-momenta (data), and 
information describing the events as one entity, an event sample (metadata). 
The latter is also not very strictly defined. For example, if the number 
of particles in an event is the same for the whole sample, we can add the 
parameter to metadata. But if it varies from event to event it is certainly 
a part of data. In our definition of metadata we try to single out the most 
common characteristic of event samples, which could be applied in most cases. 
Metadata form the main contents of MCDB. In this sense, MCDB can hold a path 
to an event sample only and the sample itself can be located somewhere else. 

The benefit of the separation is the following. MCDB interfaces provide 
the means to manipulate with metadata only\footnote{Except for interfaces 
which are responsible for downloading and uploading of event files. See 
more details in the next sections.}. This simplifies the structure and 
software of MCDB drastically. Thus by means of MCDB interfaces an end-user 
can search for a necessary event sample (according to given criteria), 
comprehend what the sample holds, and how the events were prepared. In 
other words, MCDB should let the end-user know how to reproduce the events. 
According to the idea the metadata must describe the corresponding event 
sample in a comprehensive manner. This information should be entered by the 
event sample author. In some cases, metadata are encoded inside the event 
file itself and can be inserted to MCDB (semi-)automatically. 

Comprehensive description of an event sample requires a lot of information, 
which should be entered to the database. However, in practice, in this 
specific data domain a large part of the information is common for lots of 
samples. For example, the Standard Model processes are described by a large 
set of SM couplings and particle parameters (masses, widths, etc.)\footnote{
A worse situation arises in Beyond SM models, where we can have hundreds 
of physical parameters in some models}, but usually, only few parameters 
are modified from one sample to another. In the MCDB conception we 
introduce ``Model'' -- a set of parameters, which can be attached to an 
event sample. An author of events can choose an appropriate model and 
change a few parameters in the model and store the modified set of 
parameters as a new model with a new name. The same solution is used in the 
description of MC generators. We introduce a standard record to describe 
the programs. The author simply chooses one of the standard records and 
attaches it to a new article. In order to include the features in the 
author interface we developed our own Content Management System with a 
flexible structure, which can be extended in a simple manner. 

The second idea behind the current design of MCDB is that MCDB is an area 
for interaction between two different communities, producers of events and 
consumers of the events. We call the groups ``authors'' and end-users 
respectively. Since the goals and tasks of the two groups are different, 
the corresponding MCDB interfaces intended for each group of users should 
also be different. Any researcher who feels his/her sample is worthy to come 
within MCDB can make a request to open a new author account on the MCDB server. 
It means MCDB does not assume to have a special team (of event producers) 
to prepare events according to end-user requests. 

There are several blocks in LCG MCDB, which should be realized: 
\begin{itemize}
\item 
Content Management System with a powerful and flexible Web interface for 
authors of event samples. It should have several types of templates to 
simplify the task of event sample description. 

\item 
A block of tree graph of physical categories with articles published by 
authors. This is the main part MCDB visible via Web browsers with no 
authentication in MCDB. 

\item 
A powerful search engine based on SQL/XML to search for contents of MCDB. 

\item 
A programming interface to CASTOR~\cite{Baud:2003ys}, which is used as a 
native storage of event samples. 

\item 
A block of direct uploading of files from Web/AFS/CASTOR/GRID to MCDB. 

\item 
Block of direct downloading of files from MCDB via Web/CASTOR/GRID (URI). 

\item 
A flexible and reliable authentication system based on CERN AFS/Kerberos 
logins or LCG GRID certificates

\item 
Backup system for all stored samples and corresponding SQL information. 
g
\item 
API to the LHC collaboration software environments. 

\item 
The standard record of an event sample. The record should be encoded to 
a set of SQL tables. 

\item 
A unified and flexible format of event files based on the LHEF agreement 
and the HepML language. A programming package which supports the format. 

\end{itemize}

\subsection{Metadata in LCG MCDB} 
\label{parameters}

In general, metadata can hold very specific information and can be presented 
in an arbitrary form. In fact, it is one of the main problems of knowledgebases, 
since the arbitrariness results in problems in introducing effective and 
relevant search methods in knowledgebases. Our situation is simpler than the 
general case and we can limit ourselves by some general set of parameters 
which cover most parts of the necessary information on event samples. 
Owing to specific purposes and application area of MCDB, we can define 
the standard record for MCDB articles. Now the record corresponds to a 
set of parameters stored as a record in our relational DB and a comment 
written by the author in free form. All information which is not kept within 
the standard record can be and should be put in the comment. 
MCDB search requests use the standard records to retrieve information. 
If MCDB users, authors or end-users, request to add new parameters to the 
standard record it can be extended. 

The standard information to describe event samples can be divided into 
several blocks. Each of the blocks corresponds to a definite set of 
parameters which are necessary to interpret a concrete event sample. 
The list below gives a short description of the main blocks: 
\begin{itemize}
  \item General information about a simulated event sample or a group of samples
    \begin{itemize}
      \item Title of physical process
      \item Physical Category (e.g. Higgs, Top physics or W+jets processes) 
      \item Abstract (short description)
      \item List of authors
      \item Name of an experiment and/or a group (for which the sample was prepared or intended)
      \item Author comment on the sample (some additional unstructured information on the sample)
    \end{itemize}
  \item Physical process 
    \begin{itemize}
      \item Initial state (names of beam particle, energy, etc.)
      \item Final state (name of the final particles, etc.)
      \item QCD scale(s)
      \item Process PDF (parton distribution functions) applied
      \item Information on separate subprocesses, if they are distinguished 
    \end{itemize}
  \item Event file 
    \begin{itemize}
      \item File name 
      \item The number of events 
      \item Cross section and cross section error(s) 
      \item Author comment 
    \end{itemize}
  \item Used MC generator 
    \begin{itemize}
      \item Name and version 
      \item Short description 
      \item Home page Web-address 
    \end{itemize}
  \item Theoretical model used to simulate the events 
    \begin{itemize}
      \item Name 
      \item Short Description 
      \item A set of physical parameters and their values with the author's descriptions 
     \end{itemize}
  \item Applied cuts 
\end{itemize}

%%%%%%%%%%%%%%%%%%%%%%%%%%%%%%%%%%%%%%%%%%%%%%%%%%%%%%%%%%%%%%%%%%%%%%%%%%%%%%%%%%%%%%%%%%%%%%%%
\section{LCG MCDB Software Description}
\label{soft}

This section describes shortly all subsystems and software technologies 
adopted in LCG MCDB. The current version of LCG MCDB is based on the following 
technologies: WWW, CGI, Perl, SQL, XML, CASTOR, and GRID. MCDB is a Web 
server written as a set of Perl CGI scripts. The scripts interact with 
relational DB by means of SQL requests and can generate either Web pages or 
XML documents. The main storage of event files is based on tape robots at 
CERN available via CASTOR. The MCDB software is organized as a set of 
Perl modules with the possibility of installing and customizing the software 
on other sites. All of the MCDB software has been developed from scratch 
and is available publicly in LCG CVS~\cite{lcg_mcdb_cvs_savanna}. 

\begin{figure}
\begin{center}
\epsfysize=10.0cm
\epsffile{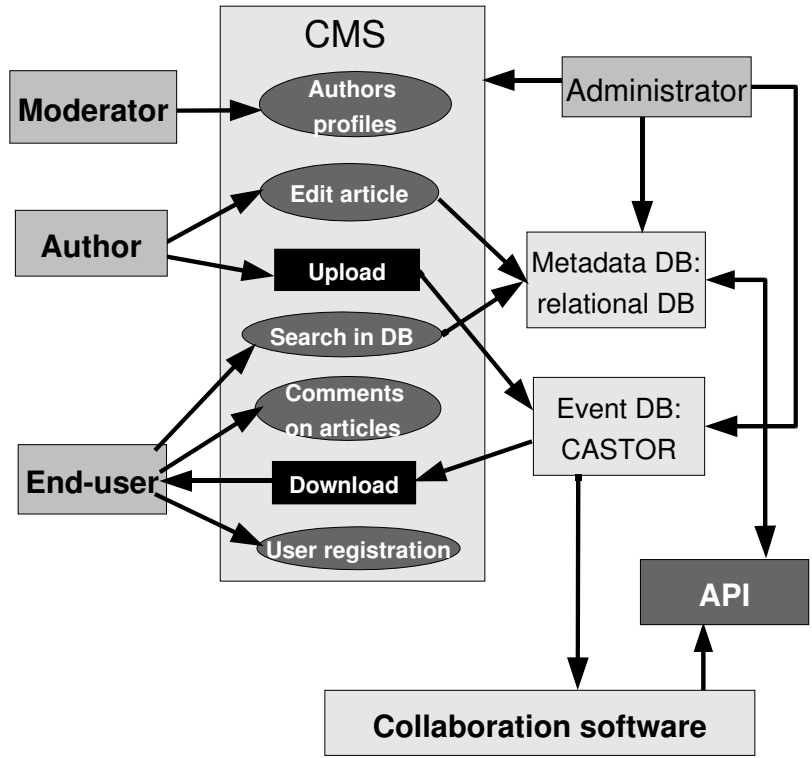}
\caption{
This general scheme of LCG MCDB shows main interfaces and interaction with 
main consumers
}
\label{lcg_mcdb_scheme}
\end{center}
\end{figure}

For the whole contents of LCG MCDB we provide a daily backup of the SQL DB 
and double mirroring of the samples in CASTOR. The main unit of MCDB is an 
{\bf article}, a document describing one or several event samples. 
MCDB articles are distributed into {\bf categories}, 
i.e. a set of articles concerning a particular type of physical process 
(e.g. top physics, Higgs physics) or theoretical model (e.g. supersymmetry, 
extra dimensions). Each category has its own branch in the main MCDB Web 
tree graph. The access system in MCDB reminds of a classical system used in 
the usual Internet forums or newsgroups. There are four different types of 
permissions to access MCDB. The {\bf end-user} access is reserved for users 
who are interested in requesting a new event sample or in downloading or 
making comments to already published event samples. The {\bf author} access 
is reserved for authorized users (MC experts) and requires registration on 
the main Web site. Only authors can upload and describe new event samples. 
The {\bf moderator} access is reserved for users who manage author profiles 
and are responsible for general monitoring of information uploaded. The 
{\bf administrator} access is reserved for software developers and maintainers 
who take care of the database itself. The scheme of LCG MCDB is shown in 
Fig.~\ref{lcg_mcdb_scheme}. 

\subsection{Web interface}
The Web-interface consists of two parts: 
\begin{itemize}
\item 
End-user area, where any user can search for a necessary event sample in 
the whole set of available samples. Requests can be done either via a search 
form or by browsing the main tree graph of categories where all articles 
are available. Users can read the description of the events, download the 
samples, ask questions about the samples and read the previous discussions 
on the particular event sample. 

\item 
Author area, where authors can upload new event files to the database and 
describe the events using the MCDB template system. As we mentioned, the 
users do not need to enter all the necessary information 
from scratch, since the templates have a lot of pre-entered information. 
Authors can interact with the end-users of their samples on public forums 
attached to each article. With the same interface, authors can edit his/her 
own previous articles or make the articles temporarily inaccessible in the 
end-user area. 

\end{itemize}

\subsection{SQL DB}
LCG MCDB adopts MySQL. The SQL technology provides the possibility to keep 
information in a very structured way. Authors provide documentation 
on event samples through forms and MCDB scripts translate the 
information to records in the MCDB relational database. 

\subsection{Search engine}
Since we use a relational DB, it is possible to provide a variety 
of complex search queries, which can use specific relations between DB 
records. The deployed Web search interface is realized as a dynamic query 
construction wizard which is based on the JavaScript XML-query constructor. 
The development of application programming interfaces to specific external 
software (for example a simulation framework of a LHC experiment) may 
benefit from similar tools in order to simplify the query construction. 

\subsection{Storage}
As a native storage interface for event samples we have selected CASTOR, 
because of the absence of serious space limitations on tapes and taking into 
account popularity of the interface in the LHC collaborations. We provide 
direct CASTOR paths for all LCG MCDB samples and also several options to 
obtain the samples through other interfaces (HTTP, GridFtp etc.). A local 
disk cache system is used to speed up the storage operations. 

\subsection{Authentication}
We pay special attention to the security of transactions during all LCG 
MCDB operations. All of the transactions are encrypted by SSL technology. 
There are two ways to log in to MCDB. The first one relies on CERN 
AFS/Kerberos login/password. The second mechanism uses LCG GRID certificates. 
Authors can choose either or both of these ways. These authentication 
methods are the standard at CERN and any CERN user can use at least one of 
these two methods. 

\subsection{Documentation}
Most of the LCG MCDB documentation is available from the dedicated web server. 
The information consists of different parts. Technical documentation describes 
the current implementation of LCG MCDB itself. The user documentation is organized 
as a set of HOW-TOs for end-users and authors. A separate documentation (available 
from the CVS repository) is intended for developers of the LCG MCDB software. 
A brief start-up manual for non-experienced LCG MCDB users is also available 
in the next section of this document. In addition, there are two freely 
accessible mailing lists dedicated to users and developers. Their addresses 
are available in the documentation section on the main web page of the server.

%%%%%%%%%%%%%%%%%%%%%%%%%%%%%%%%%%%%%%%%%%%%%%%%%%%%%%%%%%%%%%%%%%%%%%%%%%%%%%%%%%%%%%%%%%%%%%%%
\section{How to use LCG MCDB}
\label{howto}

An end-user who is going to look for and download events for a particular process 
can browse the MCDB categories and subcategories (the menu at the left side 
of the main LCG MCDB web page~\cite{lcg_mcdb_url}) and verify, whether an 
appropriate sample has already been generated. If this is the case, the 
end-user may read the sample article describing how the events have been 
prepared (check parameters of the theoretical model, generator name and 
generation parameters, kinematic cuts, etc.). At the bottom of the page 
there is a link to the uploaded file(s), as well as a direct CASTOR 
path to the sample. The web page also contains a link to the ``Users 
Comments'' interface, where end-users can ask questions about the sample and 
browse the previous discussions on the article. Users do not need any 
special authorization to carry out all the steps described above. The 
search engine provides different possibilities of search queries based on 
the set of main parameters of the article and samples. 

If someone wants to upload a new event sample or publish a new article in 
LCG MCDB (it means the user will become an author), (s)he should follow 
the following procedure: 
\begin{itemize}
\item 
Register as a new author. There is a link to the registration interface 
on the right side menu of the main MCDB Web page. Wait for a confirmation 
e-mail. 

\item 
Login to the LCG MCDB authors area. 

\item 
Choose the option ``Create New Article'' in the authors menu. It appears 
at the right side after authentication. 

\item 
Fill all necessary forms in the documentation template (title, generator, 
theoretical model, cuts, etc.) 

\item 
Upload event files in the ``Event Files'' sub-window. 

\item 
Tick the box ``Publish'' and click ``Preview/Save''. 

\end{itemize}

As we mentioned above, for authentication the author needs a valid CERN AFS 
login or a LCG GRID digital certificate. 

Authors can save unfinished articles in MCDB and resume to edit them later. 
Authors can edit their previous articles that are already published on the 
Web or make the articles publicly inaccessible for a while. 

The LCG MCDB team appreciates any bug reports, feedback, comments or 
suggestions for possible new implementations and improvements of the 
service (LCG MCDB).

%%%%%%%%%%%%%%%%%%%%%%%%%%%%%%%%%%%%%%%%%%%%%%%%%%%%%%%%%%%%%%%%%%%%%%%%%%%%%%%%%%%%%%%%%%%%%%%%
\section{API to collaboration software}
\label{api}

Apart from the MCDB server, LCG MCDB team provides application programming 
interfaces (APIs) specific to the simulation environments of the LHC 
collaborations. The main idea of these subsystems is to develop a set of 
routines for the collaboration software which would give a direct access 
to the LCG MCDB files during the MC production on computer farms. 

The most simple way to access event samples is to use direct WEB, CASTOR or 
GRID path to the event samples. This way does not require any special 
software developments on the side of collaboration software. This way, 
however, does not provide any possibility for automatic access to event 
sample description. This is the reason we developed a more complicated 
interface which could be used for automatic processing of event samples 
and the corresponding documentation. According to our idea, MCDB team 
provides API based on XML representation of event sample metadata. 
The current version of the API is a C++ library, which can be added 
to collaboration software. The XML output from LCG MCDB is based on the 
HepML~\cite{hepml},\cite{cedar} specifications (for more details, see the 
next section). 

The current library contains C++ classes and provides routines to fill the 
class objects with information from a MCDB article, including CASTOR/GRID/HTTP 
paths to event files attached to the article. Such an interface has already 
been implemented in the CMS collaboration software environment. 
The software and documentation are available in LCG CVS~\cite{lcg_mcdb_cvs_savanna}.

\begin{figure}
\begin{center}
\epsfxsize=12.5cm
\epsffile{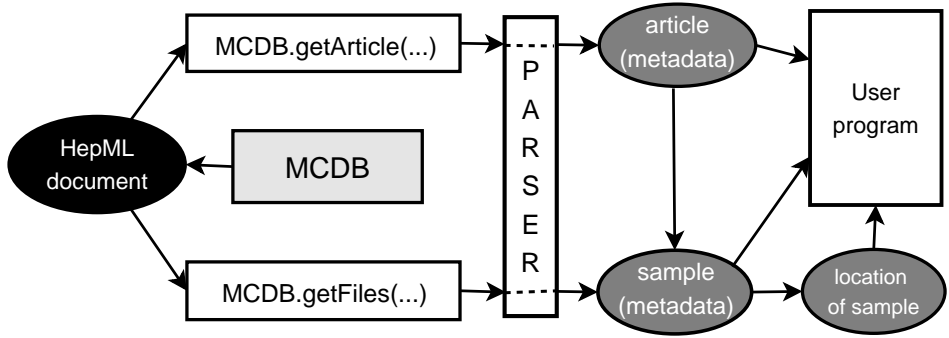}
\caption{
This general scheme shows how the API interacts with LCG MCDB to external user software.
}
\label{lcg_api_scheme}
\end{center}
\end{figure}
Fig.~\ref{lcg_api_scheme} reports a general scheme of an interaction between 
LCG MCDB to external user software via the API. 

In the future, some emphasis will be put on the development of extensions 
of API specific to the automatic uploading of HepML information and event 
samples to MCDB. This development will be carried out in the context of 
the HepML project. 

%%%%%%%%%%%%%%%%%%%%%%%%%%%%%%%%%%%%%%%%%%%%%%%%%%%%%%%%%%%%%%%%%%%%%%%%%%%%%%%%%%%%%%%%%%%%%%%%
\section{A unified XML format HepML}
\label{hepmlsec}

At present, each MC generator supports its own output format of event 
files. Authors of Matrix Element tools (the term originates 
from~\cite{Boos:2001cv}) provide interface programs to pass the events of 
a particular MC generator to the subsequent level of simulation 
(i.e. showering, hadronization, decays, simulation of detector response). 
The first step to standardize such interfaces has been described in the 
agreement ``Les Houches Accord Number One'' (LHA-I)~\cite{Boos:2001cv}, 
where a definite and strict structure of FORTRAN COMMON BLOCKS to transfer 
the necessary information from one code to another was fixed. The second 
step in this direction has been done in the agreement ``Les Houches Event File'' 
(LHEF)~\cite{Alwall:2006yp}, where the information fixed in LHA-I is translated 
to the event file structure. All other information can be kept in a specific 
place inside the header of the event file. The standard does not apply any 
limitations on the extra information and the structure of the block. The 
next natural step is to provide a unified format to keep the necessary 
information within the LHEF structure. In this context the other information 
means the metadata described in the section~\ref{concept} and some other 
information specific to the sample (parameters of matching in different 
schemes, information on specific NLO approximations, jet parameters, etc.). 

Owing to the highly dissimilar nature of the information, the most appropriate 
technology for the unified representation could be XML-based format. In 
this case it can provide the possibility to describe the stored information 
in a very flexible and structured way. 

The main idea behind the XML-based format is the flexibility to build and 
include a set of necessary parameters in an event file. For example, 
different MC generators may use the same tags for description of the 
physical parameters or they may need to keep specific information (through 
introduction of new dedicated tags). The new tags do not spoil the event 
file format and we do not need to re-write our routines which process 
these event files automatically. HepML is now being developed within 
the special LCG HepML project in collaboration with the CEDAR 
project~\cite{Butterworth:2004mu}. More information is available on our 
wiki~\cite{hepml-wiki}.

As the first part of HepML we have prepared several XML Schemas. The main 
goal of the Schemas is to provide a general and formal description of event 
data structures which are kept in XML files. Adapting the idea authors of 
MC codes can use powerful XML tools in developing of I/O routines. If the 
routines are consistent with the Schemas, event files generated by the 
routines can be read by other programs without changes in input routines of 
the programs. Also the Schemas can be used for validation of event files if 
the files are written according to HepML specifications. Now we have three 
main Schemas. The first XML Schema {\it lha1.xsd} corresponds to the whole 
set of parameters composing the LHA-I agreement. The other two Schemas, 
{\it sample-description.xsd} and {\it mcdb-article.xsd}, describe parameters, 
which are necessary to generate an XML data for an event sample and to form 
an LCG MCDB article for the sample. It means it includes all parameters 
mentioned in the list in Sect.~\ref{parameters} except for 
some parameters from LHA-I. 
The CEDAR team develops other XML Schemas for other tasks arising in the 
problem of automatization of data processing in HEP. Now all the Schemas 
are unified in one general formal XML Schema {\it hepml.xsd}, which includes 
all the other Schemas as sub-Schemas. This solution leaves freedom to 
develop Schemas and software in independent groups, but to use Schemas of 
both groups in one software project. All the developed Schemas are available 
in~\cite{hemplschemas}.

Possible internal adaption of LHEF and HepML formats into the most popular 
MC generator projects would result in a significant improvement of the MC 
event sample documentation and book-keeping. Such adaption of the unified 
standards provides the possibility to develop new standard interfaces and 
utilities. The LCG MCDB project already implements a part of HepML specifications 
in MCDB API. A dedicated document discussing the details of the requirements and 
describing the HepML proposal will appear in the near future~\cite{hepmlpaper}. 

%%%%%%%%%%%%%%%%%%%%%%%%%%%%%%%%%%%%%%%%%%%%%%%%%%%%%%%%%%%%%%%%%%%%%%%%%%%%%%%%%%%%%%%%%%%%%%%%
\section{Conclusion}
\label{conclusion}

MCDB is a special knowledgebase designed to keep event samples for the LHC 
experimental and phenomenological community. Now, a new version of the software 
has been finished and the server is ready for use by the community. Some new 
important features are implemented in the software. The features simplify and 
improve the process of documentation of event samples. 

In addition to the server, MCDB team has prepared an API for the LHC 
collaboration software environments. Implementation of the API to the 
software environments could give a possibility to use MCDB as a native 
storage in large-scale productions in collaborations. 

Subsequent development of the software will rely on further standardization 
of event file formats and elaboration of the HepML specifications and software. 

%%%%%%%%%%%%%%%%%%%%%%%%%%%%%%%%%%%%%%%%%%%%%%%%%%%%%%%%%%%%%%%%%%%%%%%%%%%%%%%%%%%%%%%%%%%%%%%%
\section{Acknowledgements}
This work was partially supported by the RFBR (the RFBR grant 07-07-00365-a). 
We thank Seyi Latunda-Dada for discussions of the text. 
We also acknowledge the LCG collaboration for support and hospitality at CERN. 
Participation of A.~S. in the project was partly supported by the 
UK Particle Physics and Astronomy Council. 

%%%%%%%%%%%%%%%%%%%%%%%%%%%%%%%%%%%%%%%%%%%%%%%%%%%%%%%%%%%%%%%%%%%%%%%%%%%%%%%%%%%%%%%%%%%%%%%%
%%%%%%%%%%%%%%%%%%%%%%%%%%%%%%%%%%%%%%%%%%%%%%%%%%%%%%%%%%%%%%%%%%%%%%%%%%%%%%%%%%%%%%%%%%%%%%%%

\end{document}